\documentclass{aa}
\usepackage{psfig,times}
\usepackage{txfonts}
\def\SS{SS 433~}
\def\xmm{XMM$-$Newton~}
\def\eta{et\,al.~}

\def\BKM{Brinkmann \eta 1991}
\def\la{~\raise.5ex\hbox{$<$}\kern-.8em\lower 1mm\hbox{$\sim$}~}
\def\ma{~\raise.5ex\hbox{$>$}\kern-.8em\lower 1mm\hbox{$\sim$}~}
\begin{document}
\title{XMM$-$Newton observations of the eastern jet of SS433}

\author{W. Brinkmann\inst{1} \and G.W. Pratt \inst{1} \and S. Rohr\inst{1}
 \and N. Kawai\inst{2} \and V. Burwitz\inst{1}  }
\offprints{W. Brinkmann}
\institute{Max--Planck--Institut f\"ur Extraterrestrische Physik,
 Giessenbachstrasse, D-85740 Garching, FRG
\and Tokyo Institute of Technology, 2-12-1 O-okayama, Meguro-ku, Tokyo
Japan 152-8551}
\date{Received ; accepted }
\abstract
{The radio supernova remnant W50 hosts at its center the peculiar
galactic X-ray binary SS\,433. It shows a central spherical structure with
two ``ears'' which are supposed to be formed
by the interaction of the precessing jets of \SS with the supernova
shell.}
{A study of the spectral and spatial structure of the X-ray emission
can reveal the physical conditions of the interaction of the precessing jet
of \SS with the material of the surrounding supernova remnant W50.}
{In two pointings in September/October 2004 for 30\,ks each the eastern jet 
of SS\,433  was observed with XMM-Newton to study the outermost parts of the
``ear'' and the X-ray bright emission region about 35\arcmin~
from SS\,433.} 
{The spectra consist of two components: a non-thermal 
power law with photon index $\Gamma \sim 2.17\pm0.02$ and  a thermal component
at a typical temperature of kT $\sim$ 0.3 keV. The X-ray emission
seems to fill the whole interior region of the radio remnant W50.
The jet terminates in the eastern ``ear'' in a ring-like terminal
shock  which indicates a flow with a kind of
 hollow-cone morphology. The spatial coincidence
of X-ray and radio emission suggests 
 physical conditions similar to those found at the outer shocks 
of ordinary supernova remnants. The bright emission region closer
to  \SS radiates non-thermally in a spatially well confined
geometry at higher X-ray energies. At soft X-rays the shape of the region
 gets blurred, centered on the hard lenticular emission.
The shape of this region and the bend in the jet propagation
direction might be caused by the interaction of
a re-collimated jet with the outer, non homogeneous interstellar
matter distribution.} 
{The physical conditions leading to the re-collimation of the jet and the
 peculiar emission morphology are far from being understood  and  require
 deeper observations as well as a
 detailed modeling of the interaction of a jet with its surroundings.} 

\maketitle

\section{Introduction}

The Galactic binary \SS emits two oppositely directed jets with
velocities of $v_j \sim$ 0.26c, which precess under an angle of
$\theta = 19.8^\circ$ with a period of $P_p = 162.5$ days (Margon
1984).  The kinetic energy of the jets is enormous and might well be
in excess of $\sim 10^{40}$ erg~s$^{-1}$.  \SS is thus regarded as one
of the most prominent objects of the class of galactic Microquasars
(Mirabel \& Rodriguez, 1998).

VLA observations confirmed the distance of the object of 5 kpc and
revealed further evidence for the precessional motion of the matter of
the jets (Hjellming \& Johnston 1981).  Doppler shifted iron lines in
the spectrum of SS433 were detected by X-ray observations with the
EXOSAT satellite (Watson \eta 1986).  As the centroids of the lines
follow the precessional Doppler curve (see e.g. Margon 1984) the
X--rays are most likely originating from the base of the jets,
emitting thermal radiation at very high temperatures.  In a series of
Ginga observations from 1987 to 1991 (Kawai \eta 1989, Brinkmann \eta
1989, \BKM , Yuan \eta 1995) and later deep ASCA observations (Kawai
et al. 1994, Kotani \eta 1996) the general picture of hydrodynamically
out-flowing cooling jets was confirmed.  Recent observations with the
high spectral resolution of the {\it Chandra} transmission gratings
(Marshall \eta 2002, Namiki \eta 2003) and the high sensitivity and
broad energy band of XMM-Newton (Brinkmann \eta 2005) have put more
stringent constraints on the physical parameters of the jets and on
the geometrical conditions of the emission region.

The general picture is that the matter at the base of the out-flowing 
jet is hot, kT$\sim$ 20\,keV,  and cools to low temperatures over a
distance of typically a few$\times 10^{12}$ cm, giving further out rise to the
observed optical line emission.  The emitted X-ray radiation, with
a luminosity of $\sim 3\times10^{35}$ erg\,s$^{-1}$, is ``driven" 
by the thermal energy of the out-flowing gas.   The much higher
(by a factor of $\la 10^5$) kinetic luminosity of the jets remains
invisible and feeds the surrounding supernova remnant W50.

Hydrodynamic models based on this 'canonical' jet model generally
provide good fits to the data but might have to be refined to
accommodate new recent observations.
Utilizing the high spatial resolution of the 
Chandra ACIS instrument Doppler shifted iron lines were found
at distances of $\ma 10^{17}$ cm from the central source 
(Migliari \eta 2002). The required in-situ re-heating of the 
emitting gas and the temporal variability of the arcsec-scale 
X-ray jet during the multiple Chandra observations might be explained
by a second, faster outflow in the jet (Migliari \eta 2005).
Noticeable variations of the Doppler shifts of the jets' emission lines
on time scales less than a day might be caused by changes in the
jets' terminal velocities which are affected by environmental effects that
disturb the direction or the speed of the jet (Marshall \eta 2005).

\begin{figure*}
%psfig{figure=jet1and2-neu-A4.ps,height=9.99truecm,width=15.0truecm,angle=0}
{\hskip 1.2cm \psfig{figure=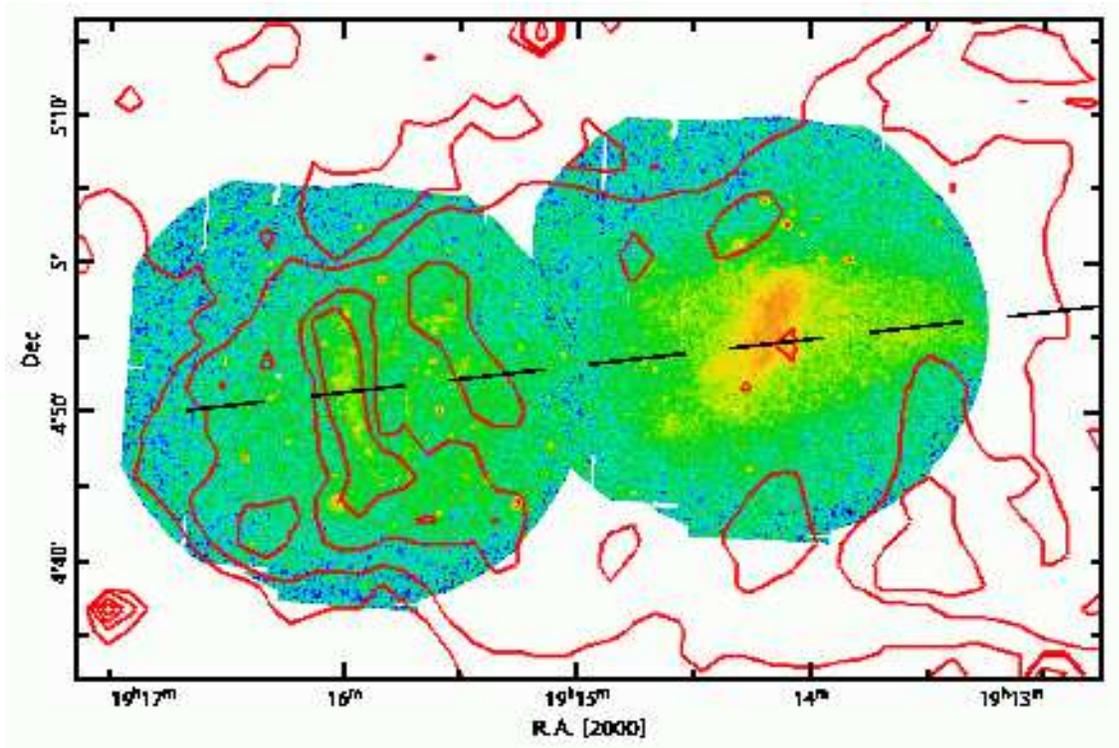,height=9.99truecm,width=15.0truecm,angle=0}}
\caption[]{\small Background and exposure corrected, smoothed  image of the
merged  PN + MOS observations of the eastern jet in the 0.3$-$10 keV energy band
superimposed on the VLA radio contours of the outer boundaries of the 
eastern ``ear'' of W50. The dashed line represents the approximate
locus from the center of the terminal shock to the position of SS\,433. }
\label{fig:totima}
\end{figure*}

At larger distances from the compact object the matter flow of the
jets interacts with the circumstellar medium and loses its energy and
momentum.  The radio shape of W50 shows two well-known lobes, the
so-called ``ears'', caused by the out-flowing jets, which seem to be
confined to the shell of the host supernova remnant W50.  The
out-flowing matter, however, can be seen only very close to SS433, on
scales of $\la 3\arcsec$ (Blundell \& Bowler 2004) and further out,
at the ``ears''.  In between, there are no radio contours indicating the
presence of a directed, collimated flow.  X--ray observations with the
Einstein IPC detector (Watson \eta 1983) demonstrated that 
the energy requirements and the morphology of the lobes are
consistent with steady beaming over the life time of the remnant
with the precessing jets supplying the directed energy. 
The emission appears
patchy, peaks at distances of $\sim$ 35 arcmin from the source and
shows hardly any flux close to SS\,433. The main contribution to the
flux comes from the narrow cone of the jets of about 20 degrees
opening angle, but there is, at a lower level, a much wider component
suggesting that energy is "leaking out" from the well confined jets.
The soft (0.5 - 4.5 keV) X-ray luminosity from each lobe is $\sim
6\times 10^{34}$ ergs~s$^{-1}$ and their spectra are considerably
softer than SS433 itself. However, the IPC data did not allow the
distinction between a thermal or a non-thermal origin of the emission.

ASCA observations of the eastern jet showed that the spectrum is non
thermal and best fitted by a power law with photon index $\Gamma $ in
the range $\sim$ 1.4 - 2.2 (Yamauchi \eta 1994).  Further, there were
indications that the spectrum hardens towards SS\,433.

Two observations with the ROSAT PSPC (Brinkmann \eta 1996, paper\,I)
provided an X-ray image of unprecedented detail.  The morphology of
the eastern and the western jet appears to be quite different, which was
attributed to different conditions in the circumstellar medium
interacting with the jets.  For the bright emission region of the
eastern jet, about 35\arcmin ~ away from SS\,433, a power law form of the
spectrum was confirmed.  The terminal shock of the eastern jet, about
80\arcmin~ from the central source, seemed to have a thermal spectrum
with a kT $\sim$ 0.2\,keV.  However, the data quality there was rather
low and the spatial structures were partly hidden by the inner support
structure of the PSPC detector.  We have therefore utilized the high
sensitivity, high spectral resolution and wide band pass of the EPIC
cameras on board XMM-Newton to perform two pointings
on the eastern jet.

The outline of the paper is as follows. In the next section we
give an overview over the \xmm observations and describe in some
detail specific aspects of the  data analysis.
In Sec. 3  we discuss the  spectral analysis of the jets and 
in  Sec. 4 we  compare the X-ray images with the radio
structure of W50.  Then, in Sec. 5, we discuss
the implications of the results for the physical properties of
 the jets and, finally, summarize the results and open questions.
\par

\section{Observations and data analysis}
\bigskip

The eastern jet of W50 was observed by \xmm in two pointings of $\sim$
30 ks each.  On September 30, 2004 the telescopes were pointed at the
position R.A.(J2000): 19$^h$ 14$^m$ 12$^s$, Dec: 04$^\circ$ 55' 47",
i.e. the bright maximum of X-ray emission from the eastern jet about
35\arcmin ~ east of SS\,433.  The outer parts of the eastern jet at
R.A.(J2000): 19$^h$ 15$^m$ 55$^s$, Dec: 04$^\circ$ 51' 20" were observed
on October 04, 2004.  The pointings were chosen such that the entire
known X-ray emitting region (Brinkmann \eta 1995) was covered.  The
EPIC PN camera was operated in Full Window mode with a medium filter.
The MOS cameras were operated in Full Window mode, both with a medium
filter as well. Although the two RGS instruments were in operation,
the weak extended sources give insufficient signal for a meaningful
spectral analysis. 
 
The EPIC data were reprocessed  with the {\small XMMSAS}  version 6.5.
The background count rates in both observations were rather low apart
from one (two) small, shorter flares.  The data from these periods
were discarded from the spectral analysis.
For the EPIC spectral analysis we selected photons with PATTERN $\le 4$ and
FLAG$=$0 for the PN and with PATTERN $\leq 12$ and FLAG$=$0 for the MOS
cameras, both in the energy range from 300 eV to 10 keV. To correct
the event files for vignetting, we add a WEIGHT column to the event
files using the SAS task {\tt evigweight}. All subsequent science
products were extracted from this column, as described in Arnaud et
al.~(2001). 

The spectral analysis of extended structures, filling the whole field
of view of the detector, represents a major challenge because of the
difficulty to define a true background region. Secondly, in this
observation, the pointing directions of the telescope were such  
that the bright structures of the X-ray emitting region fell on the
central regions of the EPIC cameras. Given the well-known spatial
distribution of the Cu K$_\alpha$ and Ni K$_\alpha$ fluorescent lines
in the EPN, using a background region from a different part of the
detector introduces added uncertainties. 

The basic background files used for the spectral analysis are the
blank-sky files accumulated by Read \& Ponman (2003). These files were
flare-filtered and events were selected according to the same PATTERN and
FLAG criteria as the source observations, and a WEIGHT column was
added to correct for vignetting. The blank-sky backgrounds were then
recast onto the sky using the aspect information from the \SS
pointings, enabling extraction of source and background spectra from
the same region of the detector. The EPIC background is dominated by
charged-particle events above $\sim 2$ keV. The intensity of this
component can vary by typically $\pm 10\%$, and must be accounted for
by renormalisation. The background renormalisation factor for each
observation was calculated in the source-free [10-12]/[12-14] keV
(EMOS/EPN) energy band, and the WEIGHT column in each background file
was adjusted accordingly. Renormalisation factors for these
observations are $\pm 5\%$ for EMOS and $<20\%$ for EPN.

Spectra were extracted from the same region in the source and
background event files using the WEIGHT column. These spectra were
vignetting corrected, thus effective area and response files
corresponding to the on-axis position were used. These were generated
using {\tt arfgen} and {\tt rmfgen}, respectively.

\subsection{The overall structure}
  
In Fig. \ref{fig:totima}  we show the total field covered by these two 
observations in the 0.5$-$10\,keV energy range.
The data from the PN and the two MOS cameras were merged, exposure 
and vignetting corrected, and smoothed with a Gaussian filter
(details of this procedure can be found in  Pietsch \eta 2004).  
The contour lines represent the outer boundary of the radio remnant W50
obtained from the  20\,cm VLA observations of  Dubner \eta  (1998).
The data show that the interior of the remnant is filled with
weak X-ray emission; no X-ray flux is detected from outside
the radio structure. However, much deeper X-ray observations are required
to unambiguously confirm the exact spatial correlation between the outer
boundaries of the radio- and X-ray emission. 
While, as already noted previously in paper\,I, the bright X-ray emission seems
not to coincide with any radio structures and is even situated in a
region of low radio surface brightness, the region near the terminal
shock shows features arguing for a common spatial origin of the radio and 
X-ray emission.

\begin{figure}
\psfig{figure=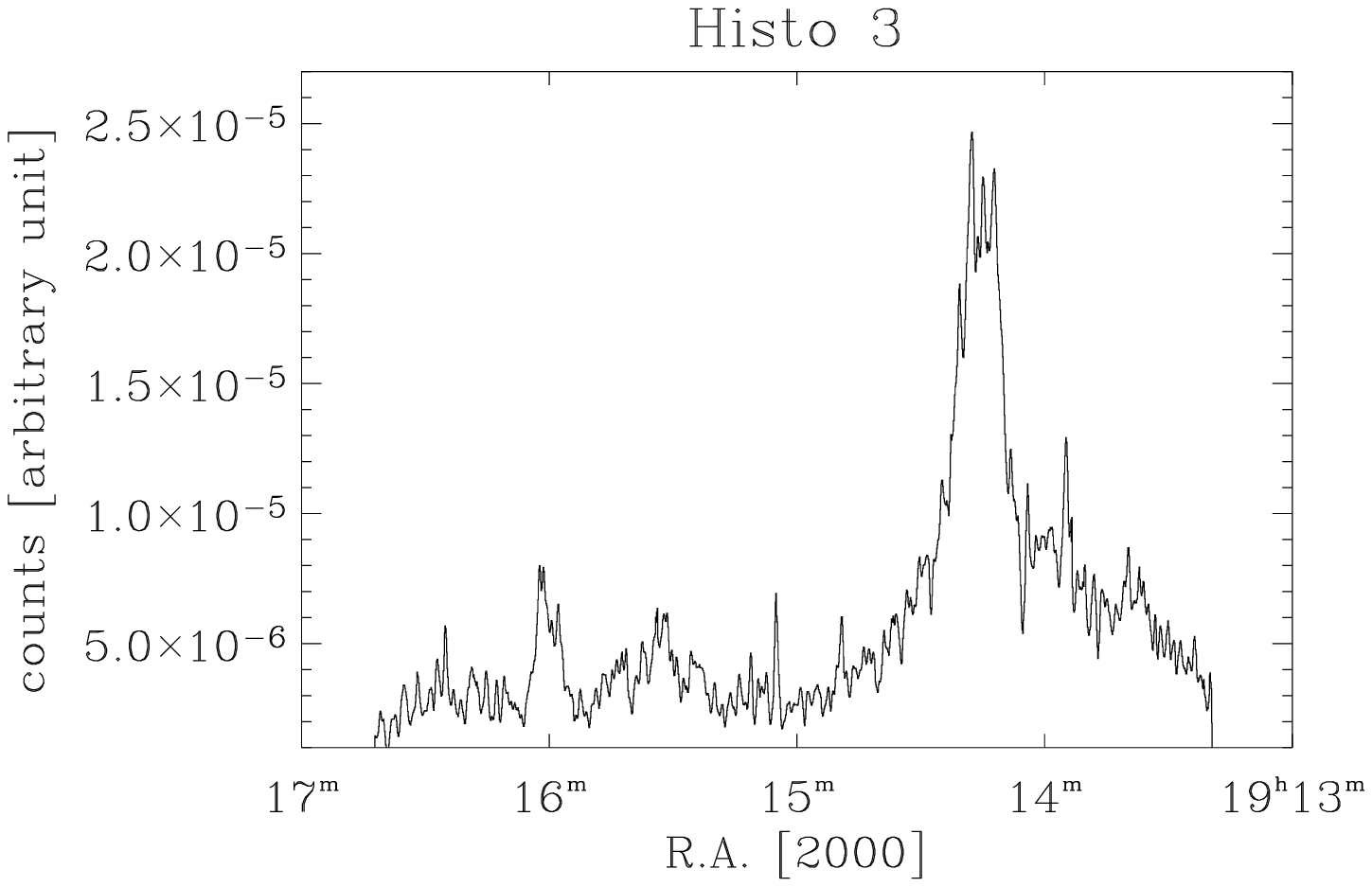,height=6.8truecm,width=8.8truecm,angle=0,%
    bbllx=70pt,bblly=94pt,bburx=465pt,bbury=346pt,clip=}
\caption[]{\small Intensity cut along the dashed axis of 
Fig.\ref{fig:totima} joining the position of SS\,433 and the center 
of the terminal shock. The maximum of the emission is about 35\arcmin~
from the position of SS\,433.
 The abscissa gives the counts/s in an image bin. } 
\label{fig:cutfig1}
\end{figure}

The dashed black line represents the locus from the centre of the
emission 
structures from the terminal jet to the position of \SS, which lies
outside of the right-hand boundary of the figure.  It forms an angle
of $\sim$ 8\degr with the line of constant declination.  In
Fig.\,\ref{fig:cutfig1} we plot the (relative) intensity of the
emission along this line.  As in the earlier
ROSAT observations 
(paper\,I) the maximum of the emission appears at about 35\arcmin~
from SS\,433. The outer structures of the terminal shock are much
weaker.  Interestingly, the intensity of emission does not start
abruptly at a certain position away from \SS (as expected from a
standing shock) but increases already along the propagation direction
of the jet before jumping to its maximum.  Note that the
figure does not provide an exact quantitative measure of the emission as raw
data over the whole energy band from the PN and the MOS cameras, which
have different energy dependent spectral responses, were merged.
 
As W50 is situated close to the galactic plane towards the galactic
center ($l \sim 40^\circ, b = -3^\circ$) we find a high background
of point like sources in the fields of view. The brightest of them are
either stars or optically unidentified and we will not further
discuss them here.  

\section{The eastern  jet}

The most prominent features in Fig. 1 are the bright emission region
at $\sim$ 19$^h$ 14$^m$ $12\fs4$, which seems to be the termination of
the jet coming into the picture from \SS (situated about 21 arcmin
outside the picture on the right-hand side), and the much weaker
fragmented structures of a 'terminal shock' in the eastern ``ear'' of W50.
The initially hot (kT $\ga$ 10 keV), thermally radiating jets emitted
from SS433 cool to temperatures below $10^5$~K on typical length
scales of $\sim 10^{12}$ cm (see \BKM ), emit the optical 'moving' 
lines at distances $\sim 10^{15}$ cm, show strong radio emission
($\ga$ 0.5~Jy at 5~GHz with spectral index $\alpha$=-0.6 ) on scales
from 10$^{14} - 10^{17}$ cm (c.f. Vermeulen 1993, for a review ), then
remain 'invisible' until they show up in X-rays again $\sim$ 15 arcmin
from SS433, i.e., at distances $\ga 6.5\times 10^{19}$ cm.  An insight
into the physical mechanisms leading to this transfer of the kinetic
energy of the jets into radiation of different frequencies, at
different distances from the source, is certainly of great interest for our
understanding of the jet phenomenon in general.

\subsection{The bright knot}
 
\begin{figure}
{\hskip 0.25cm \psfig{figure=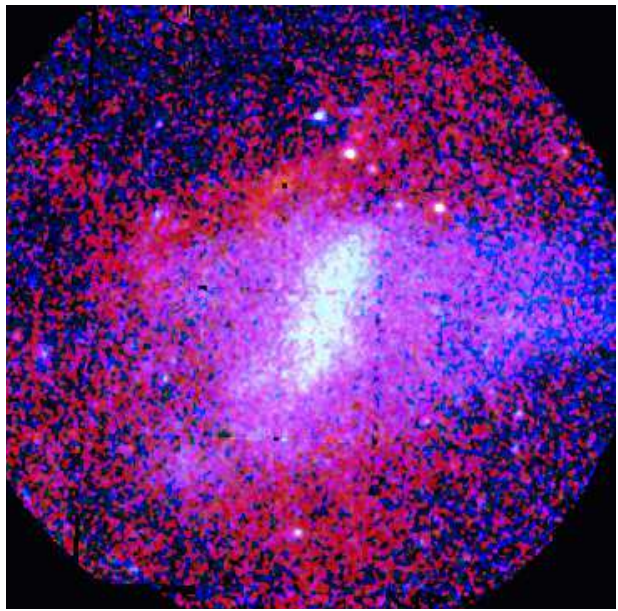,height=8.3truecm,width=8.3truecm,%
  angle=0,bbllx=127pt,bblly=235pt,bburx=503pt,bbury=590pt,clip=}}
\caption[]{\small RGB image of the bright knot from merged PN and 
 MOS images. }
\label{fig:rgb1}
\end{figure}

Using the procedures mentioned before for the production of overlaid
pictures we created images in various energy ranges and formed an
RGB - image, shown in Fig.\,\ref{fig:rgb1}. The ``red'' energy band was 
0.5$-$1\,keV, the ``green'' band 1.0$-$2.0\,keV and the ``blue''
band covered the energies 2.0$-$12\,keV. (See Fig.\,\ref{fig:jet1-s-h}
for the individual `soft' and `hard' images.)
 
The lenticular structure of the bright spot is clearly hard emission 
and this hard emission extends further to the right, towards SS\,433.
 At lower energies 
the bright spot becomes diffuse, losing its well-defined shape and 
filling a large part of the central image (red colors). At the lowest
energies the jet emission on the right side disappears.
Also visible is the clearly reduced emission region in the north-eastern
part of the image, the region outside the radio contours of W50.
The emission filling the interior of the radio remnant is rather soft;
at energies $\ga$ 2\,keV the separation  of the surface brightness
 between inside and outside W50 disappears. However, the low statistics of 
the signal at higher energies does not permit any quantitative analysis. 

\subsubsection{Spectral analysis}

For the spectral fits we first extracted for all detectors the source
photons from an ellipsoidal region coinciding with the brightest
lenticular part in Fig.\,\ref{fig:rgb1}. Background spectra were
extracted from the recast, renormalized blank-sky backgrounds as
detailed above in Sect.~2.

\begin{figure}
\psfig{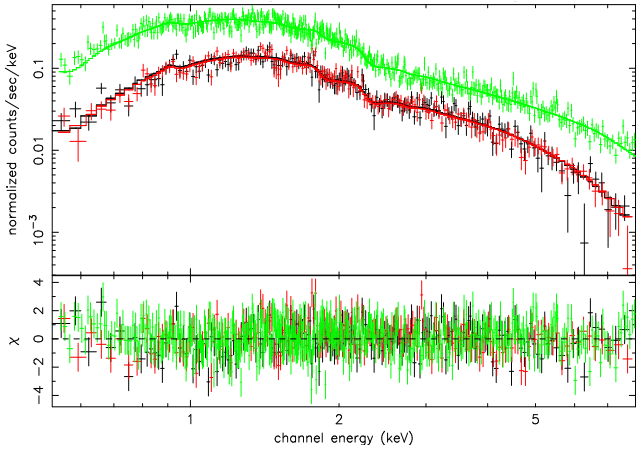}
\caption[]{Plot of the power law fit (upper panel) and residuals (lower panel)
to the  joint PN + MOS data in the $0.6-8$ keV energy band for the center of
the bright knot.}
\label{fig:pn-hard}
\end{figure}

Previous spectral analyses of this region favored a non-thermal power
law spectrum (Yamauchi \eta 1994, paper\,I).  Fig.\,\ref{fig:pn-hard}
shows the result of a simultaneous fit of the PN and MOS detectors
with an absorbed power law in the 0.6$-$8 keV energy band.  The fit
resulted in a slope of $\Gamma = 2.17\pm0.02$, an absorbing column
density of N$_{\rm H} = (0.56\pm0.01)\times10^{22}$ cm$^{-2}$ with a
reduced $\chi^2_{\rm red}$ = 1.085 for 852 d.o.f.
 However, a fit with
similar quality was achieved by a bremsstrahlung 
continuum model.  The best fit parameters were kT = 3.98$\pm$0.11 keV,
N$_{\rm H} = (0.41\pm0.01)\times10^{22}$ cm$^{-2}$ with a reduced
$\chi^2_{\rm red}$ = 1.076 (852 d.o.f.).

In both cases there were small residuals resembling weak emission
lines in the soft energy band, for example $\ma$ 0.9\,keV (Ne\,IX or
Fe\,XIX) in Fig.\,\ref{fig:pn-hard}.  We therefore repeated the fit,
first fitting a power law to the 2$-$8 keV band data only. The
extrapolation of the model showed some weak excess emission at low
energies. We then fixed the power law parameters and fitted the total
spectrum by adding a thermal ({\footnotesize MEKAL}) emission model.
The fit resulted in a $\chi^2_{\rm red}$ = 1.038 (for 853 d.o.f.) with
best fit parameters of $\Gamma = 2.41\pm0.09$, kT=0.20$\pm$0.02 keV and
the galactic absorption increased to N$_{\rm H} =
(1.07\pm0.24)\times10^{22}$ cm$^{-2}$.  The contribution of the
thermal component to the 0.2$-$2\,keV flux with $\sim
9.9\times10^{-12}$ erg\,cm$^{-2}$ s$^{-1}$ is  of the same magnitude as 
the power law component of $\sim 6.6\times10^{-12}$ erg\,cm$^{-2}$
s$^{-1}$.
 
As there were still some indications for residuals in the emission
line regions we fitted the non-equilibrium model {\footnotesize VNEI} in
{\footnotesize XSPEC}
instead of the {\footnotesize MEKAL} model in addition to the power
law.  The quality of the fit improved slightly ($\chi^2_{\rm red}$ =
1.022 (850 d.o.f.)), the temperature remained similar,
kT=0.22$\pm$0.01, and the ionization time n$_e\times t =
8.9\times10^{12}$\,s\,cm$^{-3}$ demonstrates that the plasma is nearly
in ionization equilibrium.  Leaving the elemental abundances
as free parameters resulted in values compatible with the cosmic
values inside the errors, which are quite large due to the 
limited data quality.
 
\begin{figure}
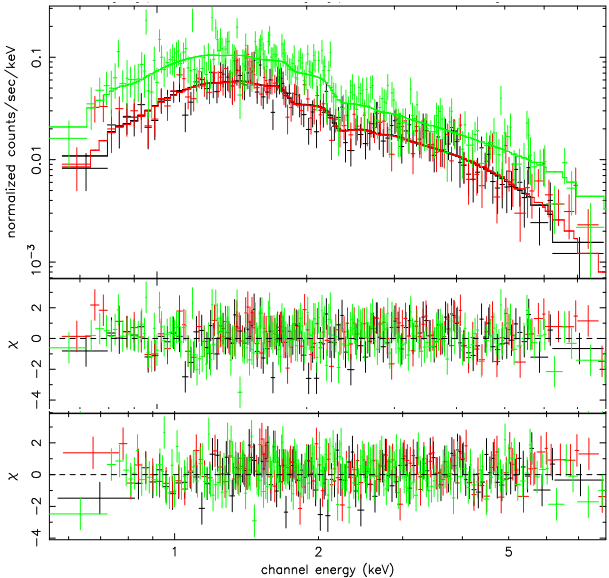

\psfig{figure=fit1_reg2.ps,height=5.4truecm,width=8.3truecm,%
  angle=-90,bbllx=112pt,bblly=30pt,bburx=514pt,bbury=711pt,clip=}
\psfig{figure=fit3_reg2.ps,height=2.4truecm,width=8.3truecm,%
  angle=-90,bbllx=383pt,bblly=30pt,bburx=570pt,bbury=711pt,clip=}
\caption[]{Plot of a power law fit (upper panel) and residuals (middle panel)
to the  joint PN + MOS data in the $0.6-8.0$ keV energy band jet region.
Bottom panel:  residuals of a power law plus {\footnotesize MEKAL} model fit
to the same data.} 
\label{fig:pos-2}
\end{figure}

To investigate the jet emission outside the pronounced lens structure
we extracted photons from a rectangular region on the jet at the right
hand boundary of the image (Fig.\,\ref{fig:rgb1}) which seems to show
only hard jet emission from the RGB colors.  Figure
\,\ref{fig:pos-2} shows in the upper two panels the power law  fit and 
the residuals to the joint PN and MOS data and the residuals.  The
power law model provides an excellent fit to the data
($\chi^2_{\rm red}$ =  0.963  for 508 d.o.f) 
with $\Gamma = 1.85\pm0.06$ and an absorbing column
density of N$_{\rm H} =(0.62\pm0.04)\times10^{22}$ cm$^{-2}$.  To
account for some smaller residuals in the soft band we first fitted
again a power law at higher energies then added an extra
{\footnotesize MEKAL} model, obtaining a
fit with nearly identical quality ($\chi^2_{\rm red}$ = 0.977 (514
d.o.f.))  over the whole energy band. The temperature was similar to
that of the fit to the 
``lens'' region (kT = 0.22$\pm$0.01) keV, the hard power law slope is
steeper ($\Gamma = 2.15\pm0.56$) compared to the simple power law fit,
and the amount of absorption increased (N$_{\rm H}
=(1.38\pm0.04)\times10^{22}$ cm$^{-2}$).  Thus, the quality of the
data does not rule out a thermal component in addition to the power
law, but does not require it either.

\subsection{The terminal shock}
 
\begin{figure}
{\hskip 0.25cm \psfig{figure=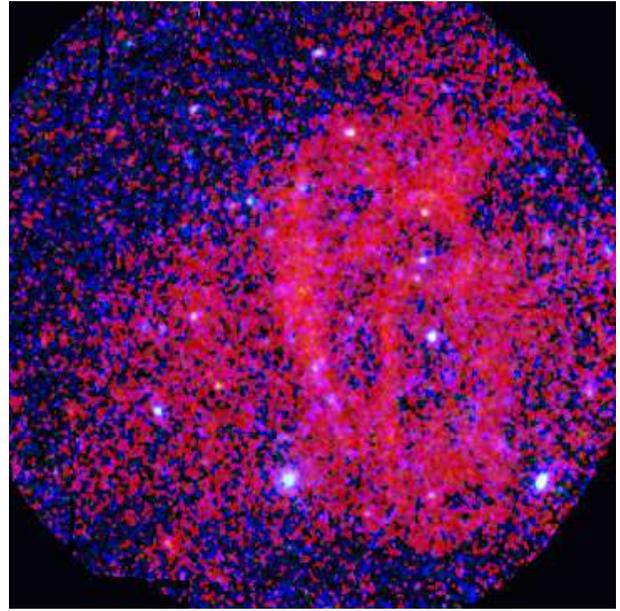,height=8.3truecm,width=8.3truecm,%
  angle=0,bbllx=127pt,bblly=235pt,bburx=503pt,bbury=590pt,clip=}}
\caption[]{\small RGB image of the terminal shock region from merged PN and
 MOS images. }
\label{fig:rgb2}
\end{figure}

Figure \,\ref{fig:rgb2} shows the RGB-image of the region of the terminal
shock.
Two features  are worth noticing: first of all the diffuse emission is
really restricted to the area inside the corresponding radio contours
of the W50 remnant (see Fig.\,\ref{fig:totima}), outside of which there is 
no emission to the level of the
background fluctuations. Secondly, the bright rim, also visible 
in the radio maps, appears to be just the front part of a 
ring-like emission structure which fills the whole circumference of the
conically converging remnant. 
Assuming that this structure is a perfect ring (doughnut) its inclination 
to the line of sight must be of the order of 30$^\circ$, not quite 
consistent with the inclination of the \SS system of $\sim 78^\circ$. 
 
The spectrum of the bright central rim is very noisy, possibly caused by the 
fact that due to its low surface brightness a relatively large 
extraction area has to be used which can include unresolved background 
sources. A simple continuum model (bremsstrahlung, power law) does not fit the
data adequately. A power law plus a  {\footnotesize MEKAL} model accounts
for some of the ``line like'' residuals, but the fit is unacceptable
($\chi^2_{\rm red}$ = 2.0 for 226 d.o.f) for the PN).
 
\begin{figure}
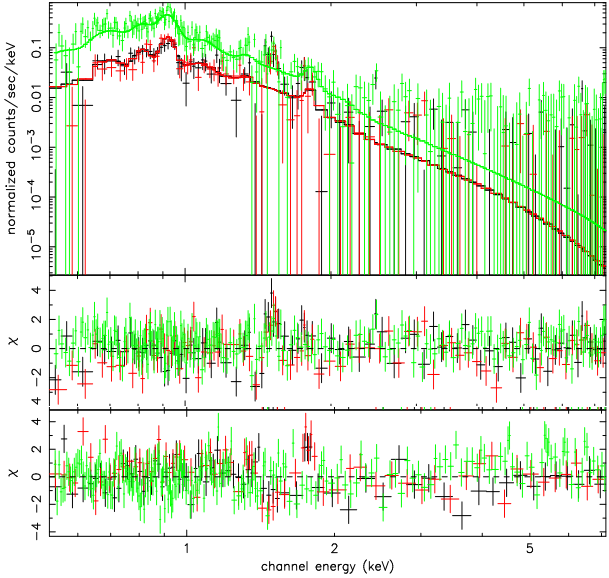

\psfig{figure=jet2-bright-pratt.ps,height=5.4truecm,width=8.3truecm,angle=-90,%
 bbllx=112pt,bblly=30pt,bburx=514pt,bbury=711pt,clip=}
\psfig{figure=jet2-bright.ps,height=2.4truecm,width=8.3truecm,angle=-90,%
 bbllx=383pt,bblly=30pt,bburx=570pt,bbury=711pt,clip=}
\caption[]{Plot of the power law  plus {\footnotesize VNEI} model fit
 (upper panel) and residuals (lower panels)
to the  joint PN + MOS data in the $0.5-7.5$ keV energy band for the 
region of the bright filament in Fig.\,\ref{fig:rgb2}.
Upper two panels: fit and residuals with blank sky background. Lower panel: 
residuals from the fit with the local background.}
\label{fig:pos-1}
\end{figure}

In Fig.\,\ref{fig:pos-1} we show the fit to the joint PN and MOS data
of the bright filament with a power law plus {\footnotesize VNEI}
model in the $0.5-7.5$ keV energy band which resulted in a
$\chi^2_{\rm red}$ = 1.07 for 648 d.o.f.  The hard power law has a slope
of $\Gamma = 4.5\pm1.1$, the temperature of the {\footnotesize VNEI}
model is kT=0.28$\pm$0.01\,keV, the ionization time n$_e\times t =
(2.9\pm1.1)\times10^{10}$\,s\,cm$^{-3}$ and the fitted absorption
N$_{\rm H} =(0.82\pm0.05)\times10^{22}$ cm$^{-2}$.  Leaving the
abundances free resulted in over-abundances of Ne, Mg, Si, Fe, and Ni,
although with values which are mostly consistent with solar abundances
within the errors.  The residuals clearly show the differences between
the various instruments, especially at low energies.
 
It is obviously hard to separate the spectrum of the ``ring'' from a
diffuse background which provides more signal than the source
itself. Another source of uncertainty is the fact that the background
is taken from a blank field accumulated from pointings at higher
galactic latitudes,  whereas the
source is seen against a background about 3.0 degrees below the
galactic plane.  We therefore repeated the fit with a local background
selected from just outside radio remnant from the current
observation, i.e. the upper left region
in Fig.\,\ref{fig:rgb2}.  This fit is relatively bad, $\chi^2_{\rm
red}$ = 1.38 (479 d.o.f.) but the best fit parameters are roughly
consistent with the above values inside their mutual errors: the power
law slope is $\Gamma = 2.29\pm0.50$, the temperature of the
{\footnotesize VNEI} model is kT=0.40$\pm$0.02\,keV, the ionization
time n$_e\times t = (2.7\pm0.3)\times10^{10}$\,s\,cm$^{-3}$ and the
fitted absorption N$_{\rm H} =(0.38\pm1.63)\times10^{22}$ cm$^{-2}$.
The fitted over-abundances are similar as above, however with formally
smaller errors which must be regarded with caution considering the
unacceptable quality of the fit.  The residuals of that fit are shown
as the bottom panel of Fig.\,\ref{fig:pos-1}.
 
A comparison of the two fits indicates that the data quality is
insufficient to accurately constrain all of the model parameters: at
higher energies  (power law slope) as well as in the soft band
(absorption) the fitted parameters depend strongly on the background
subtraction.  
   
We finally tried to determine the spectrum in the very eastern parts
of W50, to the left of the bright ring. The signal is rather low and
the fit residuals remain noisy even after all of the obvious faint
point sources have been removed from the data.  No simple model fits
the data but a power law with {\footnotesize MEKAL} model plus an
extra line at 0.924$\pm$0.07\,keV results in an acceptable fit with
$\chi^2_{\rm red}$ = 1.14  for 520 d.o.f.  The power law with $\Gamma =
1.32\pm0.13$ dominates completely the emission of the thermal
component (kT = 0.31$\pm$0.03\,keV) - apart from some prominent
emission lines.
 
\section{W50 and SS\,433}
  
The radio remnant W50 has a very distinct shape consisting of
 a nearly spherical shell centered on SS433 plus two
elongations, the  ``ears'', in the direction of the jets.
A commonly accepted model for the elongated morphology is the 
interaction between the SNR shell and the out-flowing jets.
It should be mentioned that according to our understanding of
the geometrical conditions of \SS  the jet causing the eastern 
``ear'' is pointing towards us.
Thus the X-ray spectrum emitted from the jet is a tracer of
 the physical processes leading to the slowing down of the
jet by transforming its kinetic energy into radiation.

\subsection{The morphology}
 
The X-ray brightness distribution reveals some details which are not
in agreement with the simple shell - jet interaction described above.
While the locus from \SS to the emission centroid of the
terminal shock shows good symmetry with respect to the outer X-ray and
radio emission, the lenticular structure and its axis of emission
towards \SS certainly breaks this symmetry.  Further, the angular
extent of the X-ray jet (about 18$^\circ$ as seen from SS\,433) is
much smaller than the precession angle of $\sim 40^\circ$ expected
from the kinematic model.  This indicates either some collimation of
the precessing jets (Peter \& Eichler 1993) or is a consequence of the
interaction between the jet and the SNR which generates secondary
shock waves along the symmetry axis (Vel\'azquez \& Raga 2000).

The lenticular emission structure has an angle of 20$-$30 degrees  
to the jet propagation direction, thus it cannot be a simple 
standing shock perpendicular to the jet axis. 

There is steadily increasing emission, starting outside the 
right boundary of the XMM field of view, about 15\,arcmin from \SS 
  (see Fig.\,6 of 
paper\,I), building up to a very sharp and relatively narrow peak
about 35$-$40 arcmin from \SS (see Fig.\,\ref{fig:cutfig1}). This
indicates that from distances $\ma 6\times10^{19}$ cm 
from the compact source, the kinetic energy of the out-flowing matter
is transferred to radiating relativistic electrons
at an increasing rate until an 'unstable' situation
is reached at the emission peak, where suddenly a much larger fraction
of the energy is dissipated. Further away from \SS the 
jet bends southwards towards the terminal shock, the emission 
dropping to a very low level before entering the region of the terminal
shock itself. 
 
\begin{figure}
{\hskip 0.35cm \psfig{figure=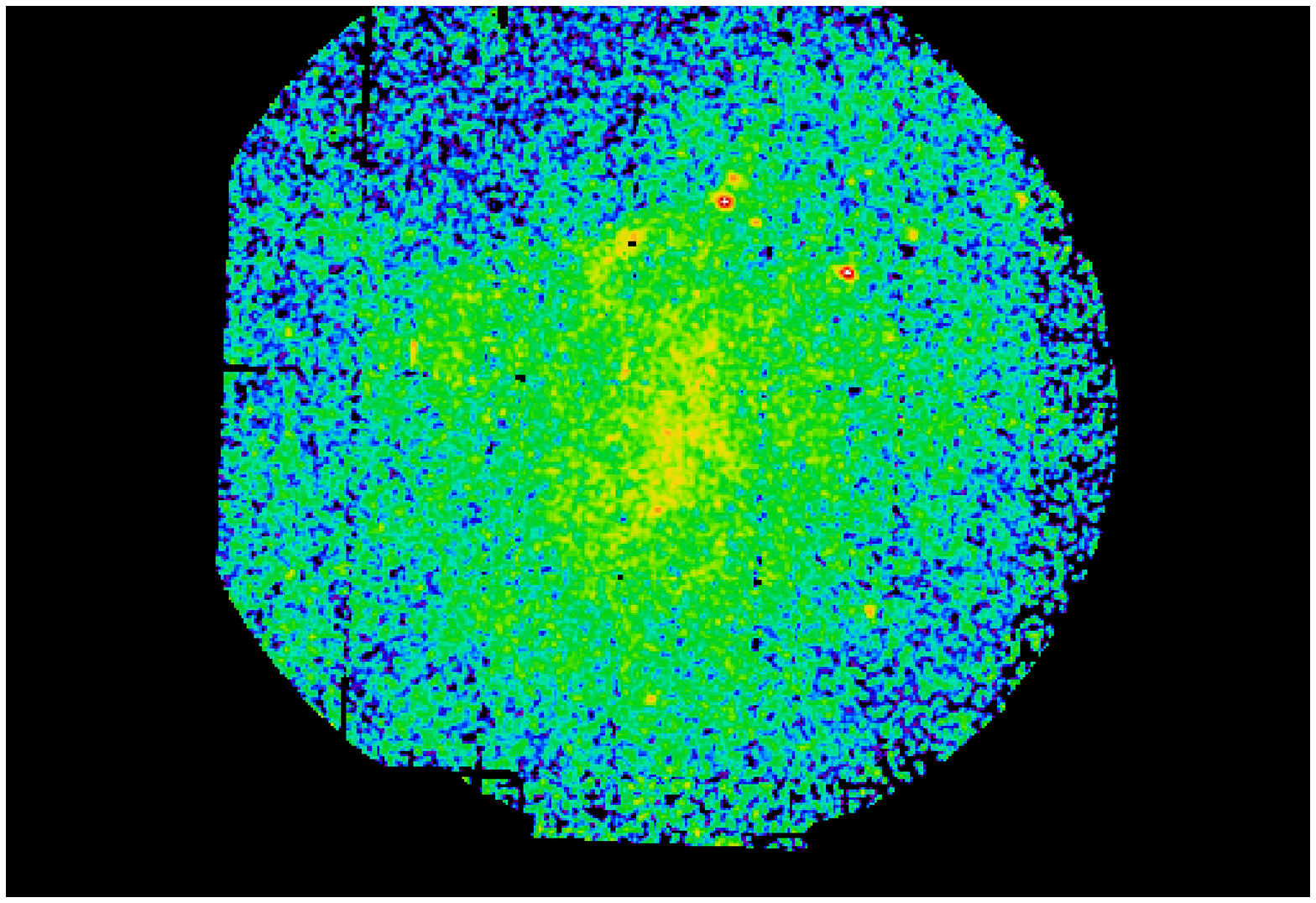,height=7.0truecm,width=8.0truecm,%
   angle=0,bbllx=185pt,bblly=296pt,bburx=475pt,bbury=549pt,clip=}}
\vskip 0.2mm
{\hskip 0.35cm \psfig{figure=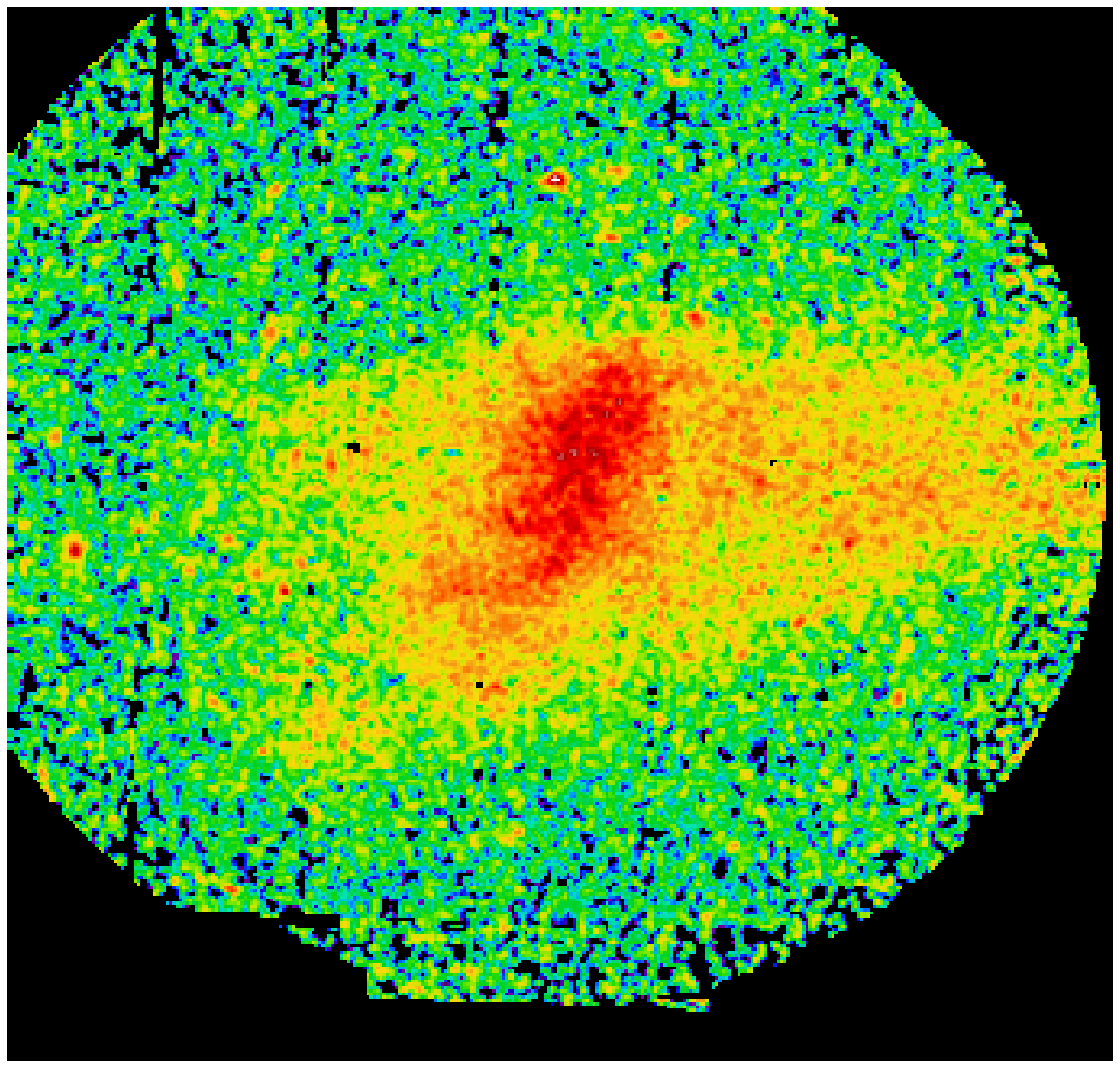,height=7.0truecm,width=8.0truecm,%
  angle=0,bbllx=185pt,bblly=296pt,bburx=475pt,bbury=549pt,clip=}}
\caption[]{\small Exposure and vignetting corrected summed PN+MOS
images of the lenticular emission region in the 0.2$-$1\,keV (top)
and 2$-$12\,keV (bottom) energy band. 
The image sizes are 8\arcmin $\times$ 7\arcmin each.  }
\label{fig:jet1-s-h}
\end{figure}
  
At higher X-ray energies the boundary of the emission zone
perpendicular to the jet direction is rather sharp, the intensity
increases typically by a factor of $\sim$ 2.5 and shows a flat
top profile. At low energies the profile smears out and widens,
as if the radiating electrons diffuse out of the acceleration zone.
This can be seen in Fig.\,\ref{fig:jet1-s-h}. In the hard 2$-$12\,keV
energy band (bottom figure) the emission region is sharply bounded 
as explained above and clearly pointing towards SS\,433. 
The soft 0.2$-$1\,keV image (top) shows only 
an extended  diffuse emission region, centered on the lenticular 
structure, without any connection to \SS on the right, outside the
images.
  
The transfer of kinetic energy of the relativistically flowing
matter into radiation is obviously a complex process and it is
not quite clear whether it is controlled `intrinsically' (i.e. 
just by the flow itself), or whether environmental conditions
play a role as well. Note that the outer shape of the surrounding 
radio remnant W50 shows  a very pronounced northern 
bay at  a similar right ascension  as the X-ray emission maximum 
($\sim$ 19$^h$ 14$^m$),
 which does not have a counterpart on the southern side. 

At the outermost eastern boundary of W50 the X-ray and the 
radio structures show an astonishingly similar appearance, 
indicating the interaction of a `terminal shock' of the jet and the
circumstellar medium.  The interior of the radio remnant 
is filled with X-ray emitting material, and the spatial coincidence
of X-ray and radio emission suggests that  the
 physical conditions  of the terminal shock region  are
 very similar to those found at the outer shocks of 'ordinary'
 supernova remnants. 
Interestingly, the prominent X-ray emission seems to form a ring-like
 structure  confined by the outer boundary of the radio remnant.
This indicates that the final jet flow might still have some kind of
 hollow-cone morphology.
  
\subsection{The X-ray spectrum}
 
XMM-Newton has for the first time the sensitivity and the wide
band pass to study in more detail the spectral behavior of the 
precessing jet interacting with its ambient medium. 
However, the photon statistics still do not allow
accurate fits of individual small spatial structures visible in the
data. Fortunately, there are no indications for drastic spectral 
changes occuring on small spatial scales (e.g., Figs 3 and 4).
 
A simple power law model provides an excellent representation of the
spectrum of the main bright lenticular area of Fig.\,\ref{fig:rgb1}.
The obtained slope of $\Gamma = 2.17\pm0.02$ is in the range
commonly found for non-thermal supernova remnants like in the Crab.
A similar power law is seen in the outer, terminal shock region
although there the low statistical significance of the high energy
data does not allow a very precise determination of the slope.
 
It must further be noticed that the bright spot can be equally well 
fitted  with a bremsstrahlung model with kT = 3.98$\pm$0.11\,keV,
a temperature in agreement with previous findings from 
ASCA (Yamauchi et al. 1994), ROSAT (paper\,I) and Einstein 
(Watson \eta 1983).
However, we tend to believe that this model is 
 merely a 'statistically acceptable' 
numerical representation of the spectral shape and not a 
correct description of the
physical emission conditions, in particular, as the data show no 
indications of 6.7\,keV iron line emission, which would be expected  at
these temperatures. 
 
In both cases, power law or simple bremsstrahlung, we find 
in the soft band spectral residuals resembling well known
emission lines. While in the lenticular spot these features
are relatively weak they are clearly dominant in the 
terminal shock region. Adding to the power law a 
{\footnotesize MEKAL} model resulted in temperatures of
$\sim 0.2-0.3$\,keV for the thermal models, but the residuals 
still showed some line features which were not accounted for
in this thermal ionization equilibrium model. 
Using the non-equilibrium model {\footnotesize VNEI} instead 
of {\footnotesize MEKAL} improved the quality of the fits 
without significant change to the basic physical parameters.
For the lenticular region the temperature remained at 
kT = 0.22$\pm$0.01\,keV, for the terminal shock we obtained
a kT = 0.28$\pm$0.01\,keV; in both cases the ionization
time scales were slightly below equilibrium ionization 
conditions and there are indications for higher than  solar 
abundances of S, Si, Fe and Ni.
 
In the fits with simple continuum models the values for the
galactic absorption turned out to be 
 N$_{\rm H} \sim 0.5\times10^{22}$ cm$^{-2}$, i.e., smaller 
than the values found for \SS itself (paper\,I, 
Brinkmann et al. 2005).  This extra absorption towards
the central source \SS might quite well be related to the
``equatorial ruff'' detected in the radio band (Blundell \eta 2001).
However, if we add a low energy thermal model to the power 
law, the fitted absorption can reach values 
N$_{\rm H} \sim 10^{22}$ cm$^{-2}$, similar to those of \SS
itself. The exact value depends on the complex interplay between the
normalization of a soft thermal model and the absorption in a 
relatively narrow soft energy band. It is further affected by the
 choice of the background
and partly on the instrument under discussion.
 
For both pointings we tried spectral fits to the regions away from the
prominent features; i.e., where the jet enters the field of view from
the right of Fig.\,\ref{fig:jet1-s-h}, and at east (left in
Fig.\,\ref{fig:rgb2}) of the ring-like terminal shock.  For the latter
the signal is very low but both a power law and thermal emission are 
required by the data.  The contribution of the power law component is
rather high which might be due to the fact that the extended source
covers the projection of the whole final outer shock region of W50.
The spectrum of the jet, entering the field of view from the right in
Fig.\,\ref{fig:jet1-s-h} is predominantly of power law origin, similar to
that of the bright lenticular region. The addition of a thermal model
does not improve the fit significantly, but cannot be ruled out
either.

\section{Physical estimates}
 
Compared to previous observations we have been able to better 
constrain the spectral parameters of the eastern jet with these  
 XMM-Newton observations.
 
Using the bright lenticular structure for the parameter estimates,
we find from the fits to the data a number density of the thermally 
radiating particles of 0.2 $\la {\rm n_e} \la $0.7 cm$^{-3}$.  
This, and the following estimates are based on the assumption of
constant density in the emission regions. Clumping of the gas will
lead to a over-estimation of the electron densities and lower
the required total mass and energy of the gas. Additionally, the total volume
of the emission region might be larger by a factor of a few than the 
chosen extraction regions for the fits.
 
Assuming cosmic abundances this thermally radiating gas has a 
cooling time scale of $\sim 10^7$ yr. The  ellipsoidal
(in 2-dimensional projection lenticular) emission region has a volume of
 $\sim 6.5\times10^{57}$ cm$^3$, containing $2\times10^{57}$ particles
with a total thermal energy content of 5$\times10^{48}$ erg.
 
The kinetic outflow energy of the jets of \SS are of the order of
10$^{39}-10^{40}$ erg\,s$^{-1}$ following early estimates and 
the numerical modeling
of the jet emission (for example \BKM, Brinkmann \& Kawai 2000). 
With its known outflow
velocity of $\sim$0.26\,c a jet thus supplies about $\la 2\times10^{44}$
particles per second into the remnant. While the energy input from
the jets over a life time of the system of $\sim 5\times10^4$ yr is 
obviously far exceeding the thermal energy content of the gas the 
number of thermal particles in the remnant must be supplied by
other means.  If the mass loss rate of the primary of the 
\SS binary system, estimated to be
$2 - 3 \times 10^{-4}$ M$_\odot$\,yr$^{-1}$ (Begelman \eta 1980,
 Fuchs \eta 2006) 
 operates over the life time of the system  a total number of
$\sim 2\times10^{59}$ particles would have been injected into
the remnant, fully sufficient to account for the density of thermal
particles in W50.
Thus it seems that the gas mass in the remnant is 
supplied by the wind from the binary system whereas the energy
comes predominantly from the out-flowing jets.

The X-ray luminosities from the outer jets represent in any
case only a small fraction of the available energy supply
by SS\,433: the 0.5$-$10 keV luminosity in the power law component of the
bright emission is L$_{pl} \ma \ 2\times10^{34}$ erg\,s$^{-1}$;
the total luminosity of the soft thermal emission might be up to a
factor of 2$-$3 higher. Even if the total emission region is 
a few times larger than the region extracted for the fits,
the total luminosity amounts only to
10$^{-5} - 10^{-4}$ of the kinetic energy of the jets.

Assuming equipartition between the energy in the magnetic field 
and the relativistic electrons, the deduced luminosity of 
the power law emission yields an estimate for the magnetic
field strength of B $\sim 2 - 4 \times 10^{-6}$ G (Watson \eta 1983,
Yamauchi \eta 1994).  The number density of the relativistic
electrons is $\sim 4\times10^{-14}$ cm$^{-3}$, the total energy 
content in the magnetic field and the electrons is 
$\sim 10^{45} - 10^{47}$ erg. 
Thus, the relativistic electrons as well represent only a small fraction 
of the kinetic energy deposited by the jets into
 the remnant over its estimated life time.
   
The most challenging physical question is how the directed
kinetic energy of the jet is transferred  to the 
relativistic electrons and then into radiation.
The simple picture of the jet being decelerated by the collision with 
the supernova shell  (Murata \& Shibazaki 1996, 
Vel\'azquez \& Raga 2000)  can hold  only for 
the terminal shock in the outermost eastern ``ear''.
 
Most of the emission, however, occurs further in, about 35\,arcmin
away from SS433 and the X-ray intensity profile along the jet axis
indicates a gradual growth of the energy transfer rate, starting at 
$\sim 6\times10^{19}$ cm from SS\,433.   If there the precessing jet would
still follow an undisturbed ballistic motion it would have
formed a conical spring with a separation of $\ma 10^{17}$ cm between the
turns which are only $\sim 1.5\times10^{15}$ cm thick. 
The radius of the ``spring'' at this
distance from \SS would be about 2.5$\times10^{19}$ cm, which is 
$\sim 5\farcm5$ at the distance of 5\,kpc for SS\,433.
 This is certainly not observed which means that
up to this distance a re-collimation of the jet must have already taken 
place. Further, early 3-dimensional hydrodynamical simulations of 
the precessing jet (Brinkmann \& M\"uller 1998) showed that already
at distances $\la 10^{17}$ cm the jet develops instabilities which lead 
to a dispersion of the matter of the ordered flow. If thus the material of the 
conical ``spring''  blends together, forming a hollow cone, and if 
most of the interior of this cone is evacuated, the jet might be 
re-collimated hydrodynamically (Peter \& Eichler 1993). If the above
discussed thermal gas fills the remnant its pressure would 
be sufficient to lead to a jet re-collimation at a distance of 
$\sim 4.9\times10^{19}$ cm (Eichler 1983) in excellent agreement
with the distance of the onset of X-ray emission from SS\,433.
 
The flow then obviously diverges again with an opening angle of $\sim 18^o$
and a steadily increasing X-ray flux. 
The process of transferring 
the kinetic energy of the jet matter to relativistic particles 
is unknown, it might be the result of a Kelvin$-$Helmholtz 
instability or some other instability related to the motion of
the jet material through the surrounding medium  (Rose \eta 1984,
Beall 1990). 
In any case the rate of energy transfer seems to grow until it
reaches a ``critical'' level, about $1.5\times10^{20}$ cm from
\SS where the X-ray emission suddenly increases by a factor of
$\sim$ 2.5, lasts for about 10$^{19}$cm and then fades away rapidly.
    
The fact that this emission structure is not perpendicular to the
jet propagation direction (as well as perhaps the sudden onset of 
the enhanced X-ray emission itself) might be related to a 
substantial gradient in the initial distribution of the interstellar 
matter to the north-east of W50: the radio shell exhibits a big ``dent''
in its outer rim, the northern part of the jet flow is braked, leading 
to a re-direction of the main flow and  an oblique lenticular
emission zone. That from here on the flow characteristic
changed markedly can be seen from the drastically reduced X-ray emission
towards the terminal shock.     

\section{Discussion and conclusions}

We have presented the data of two XMM EPIC observations of the eastern
jet of SS\,433, one pointed on the bright emission region about
 35\arcmin ~ from
SS\,433, the second on the terminal shock region in the eastern
``ear'' of W50.
The spectra can be best fitted with the superposition of two components:
a power law with photon index $\Gamma = 2.17\pm0.02 $ and a 
non-equilibrium ionization thermal component with temperatures
around kT$\sim$ 0.3\,keV. In the bright region the power law
clearly dominates, in the terminal shock region both components are
of similar strength.   
The exact strength of the thermal component is hard to determine from
the current data due to the limited data quality and `technical'
difficulties with the spectral analysis at lowest energies.

The soft X-ray emission seems to fill most of the volume of the eastern
radio remnant and no X-ray emission is seen outside the lowest
radio contours to the limit of the statistical  background
fluctuations.
 
In the inner parts of the remnant, between \SS and the bright lenticular
emission region, the jet supplies energy to its surroundings via
some unknown mechanism, obviously generating relativistic electrons
which then diffuse into larger regions of the volume of W50. 
The observed shape of the emission volume argues for an initial 
 re-focusing of the
precessing jet in the innermost $\sim 6\times10^{19}$ cm.
The energy transfer to relativistic electrons then increases, leading
to  enhanced  
emission along the jet propagation axis with an opening angle
of $\sim 20^o$. 
At a distance of $\sim 35\arcmin$ from \SS the emission brightens 
suddenly, forming at higher X-ray energies a lenticular shape,
oblique to the jet axis.   
It is proposed that this is caused by the interaction with some
exterior matter distribution which is also responsible for 
the deformation of the outer shape of the radio remnant in this region.
   
Far out in the eastern ``ear'' the X-ray emission forms a ring like 
terminal shock indicative for the braking of a hollow cone flow.
The spatial coincidence
of X-ray and radio emission suggests that  the
 physical conditions  of the terminal shock region  are
 very similar to those found at the outer shocks of 'ordinary'
 supernova remnants.
The low temperature fitted for the thermal component implies that
this terminal shock cannot be very strong, i.e., the jet flow must have
been considerably slowed down from its initial 0.26\,c. Where and how
this actually happens, remains unclear.

The total energy radiated from the jet is only a small fraction
($\la 10^{-4}$)  of the kinetic energy flux in the jets, 
integrating the synchrotron spectrum from the radio-
to the hard X-ray band.
Further, the estimated total energy 
of the inferred relativistic electrons  and the thermal gas is much smaller
(of the order of $\sim ~10^{-4}$) than
the energy injected into the remnant by the jets over its
estimated life time of $\ga 10^4$ years.
The faint extended X-ray emission from
large regions inside the remnant requires additional gas  which
can be easily supplied by the strong wind of the \SS binary system.
The injected  energy is
transported away from the sites of the interaction of the jet with its
surroundings and the sharp boundaries of the emission regions of the 
hard X-rays can provide important information about the cooling-/diffusion
time scales in the system. 
A substantial fraction of the jet's kinetic energy input might
further be consumed to accelerate the matter in the ``ears''. 
Numerical modeling is obviously 
required to estimate the efficiency of transferring the jet's
kinetic energy into relativistic particles, and to understand
where the remaining energy is going. 

Finally, another interesting result is related to SS433 itself.
For most spectral fits, \SS itself shows higher absorption than the 
X-ray lobes. 
The  excess absorption of the central object is
of the order of $\Delta N_H \sim ( 4 - 5 ) \times 10^{21}$
cm$^{-2}$.
This might be  caused by extra intervening material, e.g., from a
'cocoon' of stellar wind material blown from the primary. Such
an 'equatorial ruff' has also been detected in the radio band, 
and has been interpreted in terms of a non-isotropically emitted
outflow of matter from the binary.

\vskip 0.4cm
\begin{acknowledgements}
This work is based on observations with XMM-Newton, an ESA science mission
with instruments and contributions directly funded by ESA Member States
 and the USA (NASA). We thank Wolfgang Pietsch and Michael Bauer for their
help with the technical details of merging the images of the two observations.
\end{acknowledgements}

%\special{!userdict begin /bop-hook{gsave 300 600 translate
%20 rotate /Times-Roman findfont 80 scalefont setfont
%0 0 moveto 0.9 setgray (24.10.06) show grestore}def end}
\end{document}